\documentclass{aa}  

\usepackage{graphicx}
\usepackage{xcolor}
\usepackage{indentfirst}

\usepackage{txfonts}
\usepackage{hyperref}
\hypersetup{colorlinks=true,linkcolor=blue,citecolor=blue,filecolor=blue,urlcolor=blue}

\begin{document} 
   
   \title{Ca~{\sc ii} and H$\alpha$ flaring M dwarfs detected with multi-filter photometry}
   
   \authorrunning{P.~Mas-Buitrago et al.}
   \titlerunning{Flaring M dwarfs detected with multi-filter photometry}
 
   \author{P.~Mas-Buitrago \inst{\ref{CSIC-INTA},\ref{ucm}}
        \and J.-Y.~Zhang \inst{\ref{iac},\ref{laguna}}
        \and E.~Solano\inst{\ref{CSIC-INTA}}
        \and E.~L.~Martín \inst{\ref{iac},\ref{laguna}}
   }

   \institute{Centro de Astrobiolog\'ia (CAB), CSIC-INTA, Camino Bajo del Castillo s/n, 28692 Villanueva de la Ca\~nada, Madrid, Spain \label{CSIC-INTA} \\
   \email{pmas@cab.inta-csic.es}
        \and
            Departamento de F\'isica de la Tierra y Astrof\'isica, Facultad de Ciencias F\'isicas, Universidad Complutense de Madrid, 28040 Madrid, Spain \label{ucm}       
        \and
            Instituto de Astrof\'isica de Canarias, Calle V\'ia L\'actea s/n, 38204, San Cristobal de La Laguna, Tenerife, Spain \label{iac}
       \and
            Departamento de Astrof\'isica, Universidad de La Laguna, 38206, La Laguna, Tenerife, Spain \label{laguna}
        }

   \date{Received 26 September 2024 / Accepted 13 February 2025}

% \abstract{}{}{}{}{} 
% 5 {} token are mandatory
 
  \abstract
   {Understanding and characterising the magnetic activity of M dwarfs is of paramount importance in the search for Earth-like exoplanets orbiting them. Energetic stellar activity phenomena, such as flares or coronal mass ejections, which are common in these stars, are deeply connected with the habitability and atmospheric evolution of the surrounding exoplanets.}
   {We present a follow-up of a sample of M dwarfs with strong H$\alpha$ and Ca~{\sc ii} H and K emission lines identified with J-PLUS photometry in a previous work.}
   {We collected low-resolution NOT/ALFOSC and GTC/OSIRIS spectra, measuring the PC3 index for the spectral type determination. We used two-minute-cadence calibrated TESS light curves to identify and characterise multiple flares and to calculate the rotation period of the two active M dwarfs found in our sample.}
   {We confirm that the strong emission lines detected in the J-PLUS photometry are caused by transient flaring activity. We find clear evidence of flaring activity and periodic variability for LP 310-34 and LP 259-39, and estimated flare energies in the TESS bandpass between $7.4\times10^{30}$ and $2.2\times10^{33}$\,erg for them. We characterised LP 310-34 and LP 259-39 as very rapidly rotating M dwarfs with Ca~{\sc ii} H and K and H$\alpha$ in emission, and computed a rotation period for LP 259-39 for the first time: $P_{\rm rot}=1.69\pm0.02$\,d.}
   {This work advocates the approach of exploiting multi-filter photometric surveys to systematically identify flaring M dwarfs, especially to detect episodes of strong Ca~{\sc ii}  H and K line emission, which may have important implications for exoplanetary space weather and habitability studies. Our results reveal that common M dwarfs experience flare events in Ca~{\sc ii} H and K in addition to well known H$\alpha$ flares.}

   \keywords{stars: late-type --
                stars: low-mass --
                stars: activity --
                stars: flare --
                stars: rotation --
                techniques: spectroscopic
               }

   \maketitle

%-------------------------------------------------------------------

\section{Introduction} \label{sec:intro}

\noindent With lifespans of tens of billions of years \citep{adams1997}, M dwarfs are highly prevalent in the stellar population of the Galaxy, making up approximately 70\% of it \citep{henry1994,reid1995,reyle2021}. Thanks to their small size and low luminosity, the habitable zone of these cool stars is much closer than in their solar-like counterparts.\ This facilitates the detection of potentially habitable Earth-like exoplanets orbiting them, making M dwarfs major targets in the search for exoplanets. On the other hand, the nearby habitable zone around M dwarfs makes exoplanets more exposed to energetic events linked to stellar activity \citep{tilley2019,gunther2020,chen2021}, such as flares or coronal mass ejections. Stellar flares are sudden releases of magnetic energy caused by magnetic reconnection events in the stellar atmosphere; they are accompanied by bursts of  isotropic electromagnetic radiation \citep{benz2010} and are frequent phenomena in low-mass stars. Flares can release energies  of up to $\sim10^{37}$\,erg \citep{davenport2016} in time spans ranging from minutes to several hours, and their spectrum is often modelled as a blackbody with a temperature in the range $9\,000-10\,000$\,K \citep{davenport2020,gunther2020}.

A large fraction of M dwarfs are magnetically active, with a chromospheric activity often diagnosed using H$\alpha$ or Ca~{\sc ii} H and K line emission \citep{cincunegui2007,ibanezbustos2023}. The relation between the chromospheric emission in these lines is not straightforward; it exhibits a complex behaviour when studied over a large sample of M dwarfs because the activity level in these two regions is not always correlated \citep{meunier2024}. Moreover, flare emission in the optical domain is also known to occur in these lines \citep{heinzel1994,kowalski2013}. In optical spectra of low-mass stars, strong emission lines have been detected serendipitously, both as transient flare events \citep{Martin2001,Liebert2003,Schmidt2007} and as steady features \citep{Mould1994,Martin1999_2,Burgasser2011}.

Recent multi-filter photometric surveys, such as the Javalambre Photometric Local Universe Survey \citep[J-PLUS;][]{Cenarro2019} or the upcoming Javalambre Physics of the Accelerated Universe Astrophysical Survey \citep[J-PAS;][]{J-PAS}, with unique systems of 12 and 56 optical filters, respectively, may enable new ways to systematically detect strong emission lines in low-mass stars \citep{masbuitrago22}. The spectral energy distributions (SEDs) provided by these surveys are suitable for identifying excess emission in narrow-band filters located at specific spectral features, such as the H$\alpha$ or Ca~{\sc ii} H and K lines, which could be caused by stellar flaring activity. Although the short exposure time characteristic of multi-filter photometric surveys hinders the detection of flare events, the large number of stars that can be analysed simultaneously in this way makes it feasible.

High-precision, high-cadence photometric light curves (LCs) provided by surveys such as \textit{Kepler} \citep{kepler}, the \textit{Kepler} extended mission \citep[K2;][]{k2}, and the Transiting Exoplanet Survey Satellite \citep[TESS;][]{tess} are a powerful asset for understanding and studying the stellar activity of low-mass stars and brown dwarfs \citep[e.g.][]{martink1,doyle2019,doyle2022,kumbhakar2023}. TESS is a NASA mission launched in April 2018 with the primary objective of searching for transiting exoplanets around bright, nearby stars. Operating at optical wavelengths, the high-cadence photometric data provided by TESS capture insightful information about the magnetic activity of target stars, mainly in the form of periodic variability due to co-rotating star spots or as sudden brightness outbursts caused by stellar flares.

We present a spectroscopic follow-up of the sample of M dwarfs with strong H$\alpha$ and Ca~{\sc ii} H and K emission identified by \citet{masbuitrago22} using J-PLUS photometry. Using TESS LCs, we performed a comprehensive study of the stellar activity of two active M dwarfs found in our sample, including a detailed analysis of their flaring activity and rotation period, as well as an estimation of their age. Moreover, we discuss the possible implications of the observed stellar activity on the evolution and habitability of exoplanets orbiting low-mass stars. In Sect. \ref{sec:obs} we describe the spectroscopic observations collected for our sample. The reduced spectra of our targets and the analysis performed using TESS LCs are discussed in Sect. \ref{sec:results}. Section \ref{sec:habitability} presents a discussion of the impact of the magnetic activity of M dwarfs on planetary habitability. Finally, the main conclusions of this work are summarised in Sect. \ref{sec:conclusions}.

%--------------------------------------------------------------------

\section{Observations}\label{sec:obs}

\subsection{Sample selection}\label{sec:sample}

\noindent The sample studied in this work is the result of the search for strong emission lines performed in our previous work \citep{masbuitrago22}, using multi-filter optical photometry from J-PLUS \citep{Cenarro2019}. For this, we developed a Python algorithm capable of detecting excess in the J-PLUS filters corresponding to the H$\alpha$ ($J0660$) and Ca~{\sc ii} H and K ($J0395$) emission lines. Following this approach, we identified eight M dwarfs with emission excess in these filters (four of them in each of the filters and none showing both excesses simultaneously). In the end, one of these objects was discarded for spectroscopic follow-up because it was not bright enough, resulting in a final sample of seven M dwarfs. Table~\ref{tab:targets} lists the selected targets.

The J-PLUS SED of each target star is provided in Appendix \ref{app_a}. The excess emission in the $J0395$ filter is evident for J-PLUS0114, J-PLUS0744, J-PLUS0807, and J-PLUS0903. On the other hand, the SEDs of J-PLUS0226, J-PLUS0708, and J-PLUS0914 show strong emission in the $J0660$ filter. We attribute this behaviour to the fact that the star experiences flaring activity during the corresponding J-PLUS observing block, in which all filters are observed sequentially. The strategy for each J-PLUS observing block is to obtain, for the same pointing, three consecutive exposures per filter, with a total exposure time of approximately one hour \citep{Cenarro2019}. Flaring phenomena during the exposures for the filters of interest would explain the SED behaviour found. Given the low probability of observing a flare during the exposures for the filters of interest, it is easier to detect the less energetic and shorter-lived flares, which are more frequent and last a few minutes, as we confirm in Sect. \ref{sec:flares}.

The estimated effective temperatures for these objects, obtained with the tool \texttt{VOSA}\footnote{\url{http://svo2.cab.inta-csic.es/theory/vosa/}} \citep{vosa}, classify them as mid-M dwarfs except for one, namely LP 310-34, which has a $T_{\rm eff}=2\,500$\,K. As mentioned in \citet{masbuitrago22}, we carried out a spectroscopic follow-up for LP 310-34 that confirmed it as a late M dwarf (dM8) with H$\alpha$ in emission \citep{Schmidt2007}.

\begin{table*}
\fontsize{10pt}{10pt}\selectfont
 \caption{Targets selected for spectroscopic observation.}
 \label{tab:targets}
 \centering          
 \begin{tabular}{l l c c l c c c}
  \hline\hline
  \noalign{\smallskip}
  
  Object\,$^{(a)}$ & SIMBAD name & $\alpha\,^{(b)}$ & $\delta\,^{(b)}$ & TIC ID\,$^{(c)}$ & $T$\,$^{(c)}$ & \texttt{VOSA} $T_{\rm eff}$ & Excess \\
   & & [J2016.0] & [J2016.0] & & [mag] & [K] & \\
  
  \noalign{\smallskip}
  \hline
  \noalign{\smallskip}
  
  J-PLUS DR2 J0114+07 & \ldots & 01:14:07.54 & 07:56:32.2 & 376905949 & $15.78\pm0.01$ & 3\,200 & Ca~{\sc ii} HK\\
  
  J-PLUS DR2 J0226+34 & \ldots & 02:26:44.20 & 34:45:35.0 & 285430298 & $18.91\pm0.05$ & 3\,300 & H$\alpha$\\
  
  J-PLUS DR2 J0708+71 & \ldots &  07:08:44.51 & 71:54:25.3 & 99546413 & $18.19\pm0.01$ & 3\,200 & H$\alpha$\\

  J-PLUS DR2 J0744+40 & \ldots & 07:44:34.50 & 40:08:44.7 & 21370349 & $14.79\pm0.01$ & 3\,100 & Ca~{\sc ii} HK\\
  
  J-PLUS DR2 J0807+32 & LP 310-34 & 08:07:25.60 & 32:13:06.0 & 461654150 & $14.72\pm0.01$ & 2\,500 & Ca~{\sc ii} HK\\

  J-PLUS DR2 J0903+34 & LP 259-39 & 09:03:41.95 & 34:48:18.6 & 166597074 & $13.83\pm0.01$ & 3\,200 & Ca~{\sc ii} HK\\

  J-PLUS DR2 J0914+23 & \ldots & 09:14:05.74 & 23:52:24.9 & 85994888 & $18.39\pm0.08$ & 3\,200 & H$\alpha$\\

  \noalign{\smallskip}
  \hline
 \end{tabular}
 \tablefoot{$^{(a)}$ Hereafter we use J-PLUSHHMM as an abbreviation. $^{(b)}$ From \textit{Gaia} Data Release 3 \citep[DR3;][]{gaiadr3}. $^{(c)}$ TESS Input Catalog identifier and TESS magnitude from \citet{tic8_2}.  } 
\end{table*}

\subsection{Observational details}
\noindent We collected low-resolution optical spectra of our seven targets with the Alhambra Faint Object Spectrograph and Camera (ALFOSC) mounting on the 2.56-m Nordic Optical Telescope (NOT) with proposal number 66-208 (PI ELM). Also, we observed two bright targets (J-PLUS DR2 J0807+32 and J-PLUS DR2 J0903+34 in Table \ref{tab:targets}) with the Optical System for Imaging and low-Intermediate-Resolution Integrated Spectroscopy (OSIRIS) mounting on the 10.4 m Gran Telescopio Canarias (GTC), at the Roque de los Muchachos Observatory on the island of La Palma, Spain, with programme GTCMULTIPLE2I-22B (PI ELM).

ALFOSC is equipped with a Teledyne e2v CCD231-42-g-F61 back illuminated, deep depletion, astro multi-2 detector. The detector dimension is 2\,048$\times$2\,064 pixels with a scale of 0.2138 arcsec/pix. The NOT/ALFOSC observation was executed under visitor mode on the nights of January 26-27, 2023 (observers PMB and JYZ). We used a 1.0 arcsec slit, and \#4 grism, which provide a wavelength range from 3\,200 \AA\ to 9\,600 \AA\ with a resolution power $R\approx360$.

OSIRIS is a commonly used instrument of GTC. It covers the wavelength range $3\,650- 10\,050$\,\AA\ and has an effective field of view of 7.5$\times$6.0 arcmin. OSIRIS has two Marconi CCD44-82 (2\,048$\times$4\,096 pixels) detectors with a gap in between. The 2$\times$2 binned pixel size is 0.254 arcsec/pix. In the mode of long-slit spectroscopy, the object is centred on the slit at the coordinate $\rm{X}
=250$ of CCD2. The GTC/OSIRIS observation was executed under service mode. We required a maximum seeing of 1.2\,arcsec, a cloud free sky, and a grey moon phase, using the R1000B grism and a 1.2 arcsec slit under the parallactic angle. This configuration yields a wavelength coverage from 3\,600\,\AA\ to 7\,900\,\AA\ with a resolution power $R\approx500$. The record of observations is provided in Appendix \ref{app_b}.

%--------------------------------------------------------------------

\subsection{Data reduction}
\noindent We reduced both the GTC/OSIRIS and NOT/ALFOSC data using v1.12 of \texttt{PypeIt} \citep{pypeit:zenodo,pypeit:joss_pub}, a community-developed open-source Python package for semi-automated reduction of spectroscopical data in astronomy. \texttt{PypeIt} supports a long list of spectrographs and provides the code infrastructure to automatically process the image, identify the slit in a given detector, extract the object spectra, and perform wavelength calibration. We observed the standard stars HD~19445 and Feige~110 for flux calibration of NOT/ALFOSC and GTC/OSIRIS data, respectively.

%--------------------------------------------------------------------

\section{Results and discussion}
\label{sec:results}

\subsection{Reduced spectra}
\label{sec:reduced_sp}

\begin{figure*}
    \centering
        \includegraphics[width=8.5cm]{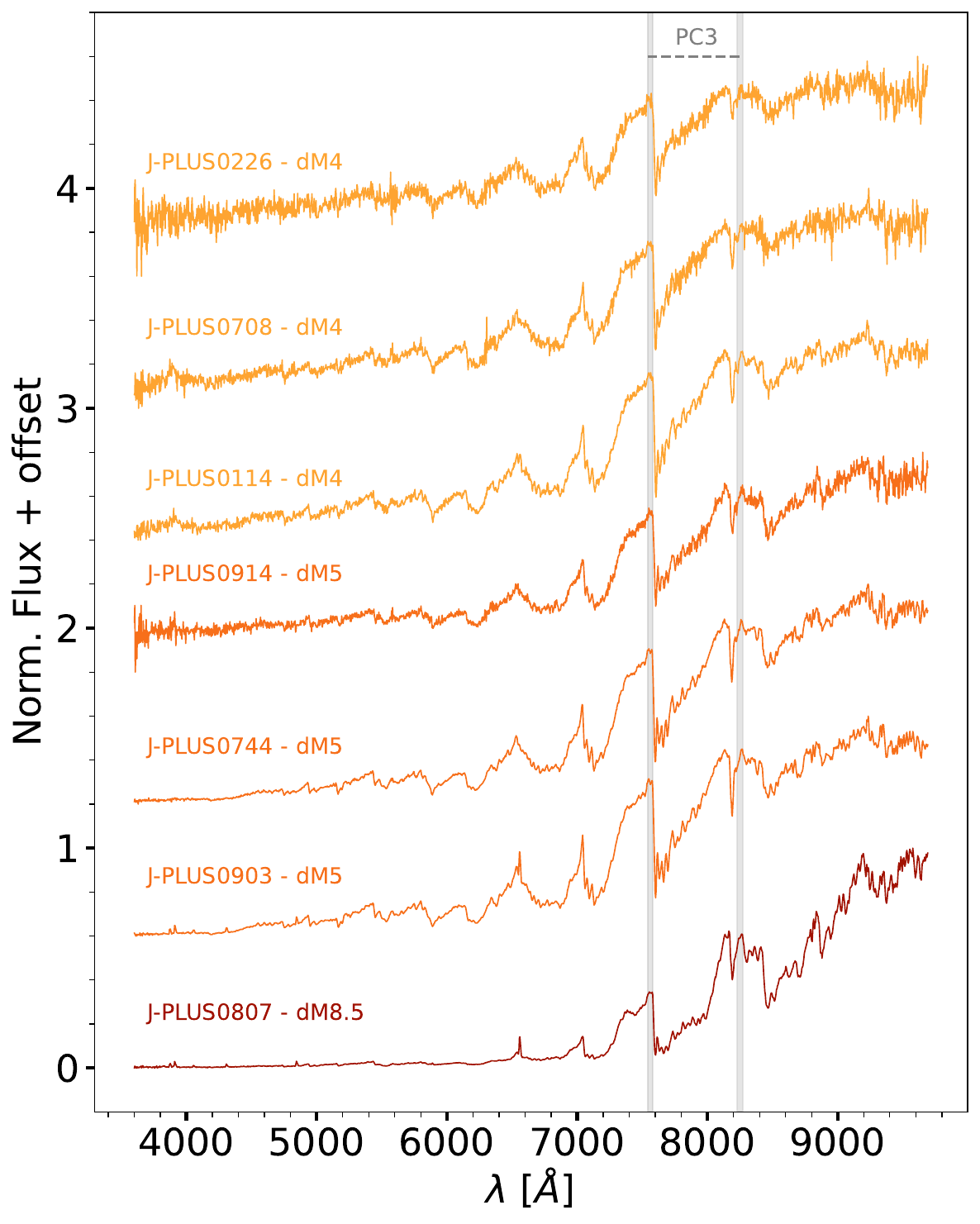}
            \includegraphics[width=8.5cm]{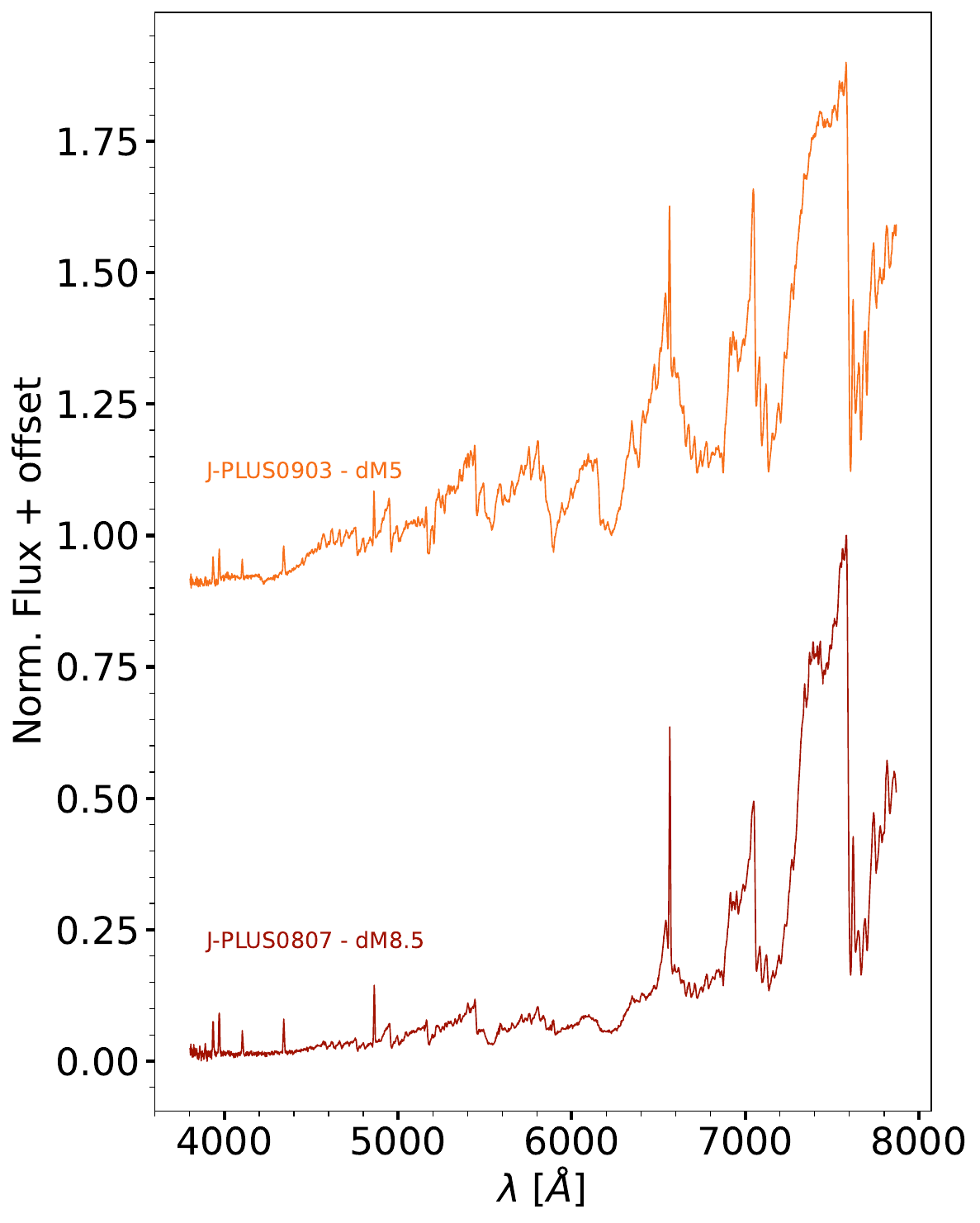}
    \caption{Co-added spectra observed with NOT/ALFOSC (\textit{left panel}) and GTC/OSIRIS (\textit{right panel}), sorted from top to bottom by the derived PC3 index (see Table \ref{tab:pc3}). The grey shaded bands in the \textit{left panel} show the spectral regions used to derive the PC3 index.}
    \label{fig:spectra_all}
\end{figure*}

\noindent Figure \ref{fig:spectra_all} shows the co-added spectra for the observed targets. We note that no intense steady line emission is observed in the spectra \citep[for examples of strong line emission in low-resolution spectra, see Fig. 1 in \citealp{Burgasser2011} or Figs. 10 and 11 in][]{Schmidt2007}, confirming that the excess emission detected in the J-PLUS photometry is not steady and is indeed caused by transient flaring activity. Moreover, objects with excess emission in the J-PLUS Ca~{\sc ii} H and K filter show no apparent differences in their spectral features compared to objects with excess emission in the J-PLUS H$\alpha$ filter (see Table \ref{tab:targets}), suggesting that the flaring activity detected in Ca~{\sc ii} H and K is not particular to a specific type of star. Hence, it follows that common M dwarfs experience two types of flares, those already well known in H$\alpha$ and those in Ca~{\sc ii} H and K revealed in this work.

Several spectroscopic indices have been explored for the spectral classification of M dwarfs \citep{lepine2003} and, in particular, for late-M dwarfs using low-resolution optical spectra \citep{kirkpatrick1995, martin1996, martin1999}. To derive spectral types for our sample, we measured the PC3 index \citep{martin1999}, which is a reliable indicator of spectral type in the [M2.5, L1] range and has been used consistently in the literature \citep{crifo2005,martin2006,martin2010,phanbao2006,reyle2006,phanbao2008}. The PC3 index is a pseudo-continuum spectral ratio between the $8230-8270$\,\AA\,(numerator) and $7540-7580$\,\AA\,(denominator) intervals, which can be used to derive spectral types between M2.5 and L1 following the calibration presented by \citet{martin1999}:

\begin{equation}    
    \rm{SpT}=-6.685+11.715\times (\rm PC3)-2.024\times (\rm PC3)^2.
    \label{eq:spt}
\end{equation}

Table \ref{tab:pc3} lists the PC3 index and the adopted spectral type, with an uncertainty of $\pm0.5$ sub-classes, for our targets. The classification obtained for J-PLUS0807 is consistent with that provided in \citet{Schmidt2007}, who derived a spectral type of dM8, with an uncertainty of $\pm0.5$ sub-classes, by visual comparison of the spectra to spectral standards. These results confirm the rest of our sample, still spectroscopically unclassified in the literature, as mid-M dwarfs.

\begin{table}
\fontsize{10pt}{10pt}\selectfont
 \caption{PC3 index and adopted spectral type for our targets.}
 \label{tab:pc3}
 \centering          
 \begin{tabular}{l c l}
  \hline\hline
  \noalign{\smallskip}
  
  Object & PC3 & SpT\\
  
  \noalign{\smallskip}
  \hline
  \noalign{\smallskip}
  
  J-PLUS0114 & 1.12 & dM4 \\
  
  J-PLUS0226 & 1.09 & dM4 \\
  
  J-PLUS0708 & 1.12 &  dM4 \\

  J-PLUS0744 & 1.29 & dM5 \\
  
  J-PLUS0807 & 1.94 & dM8.5 \\

  J-PLUS0903 & 1.29 & dM5 \\

  J-PLUS0914 & 1.29 & dM5 \\

  \noalign{\smallskip}
  \hline
 \end{tabular}
\end{table}

\begin{figure*}
    \centering
        \includegraphics[width=8.5cm]{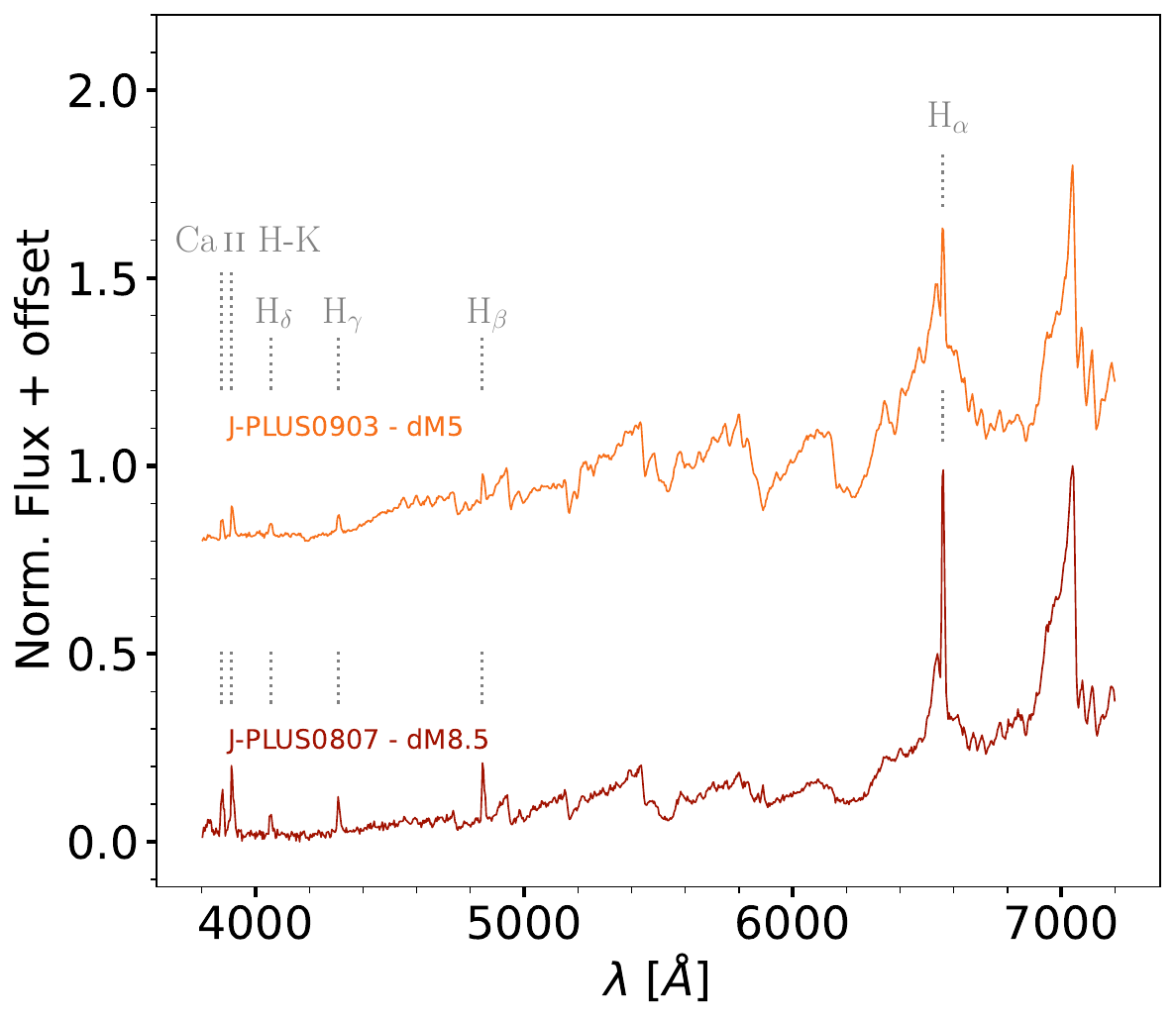}
            \includegraphics[width=8.5cm]{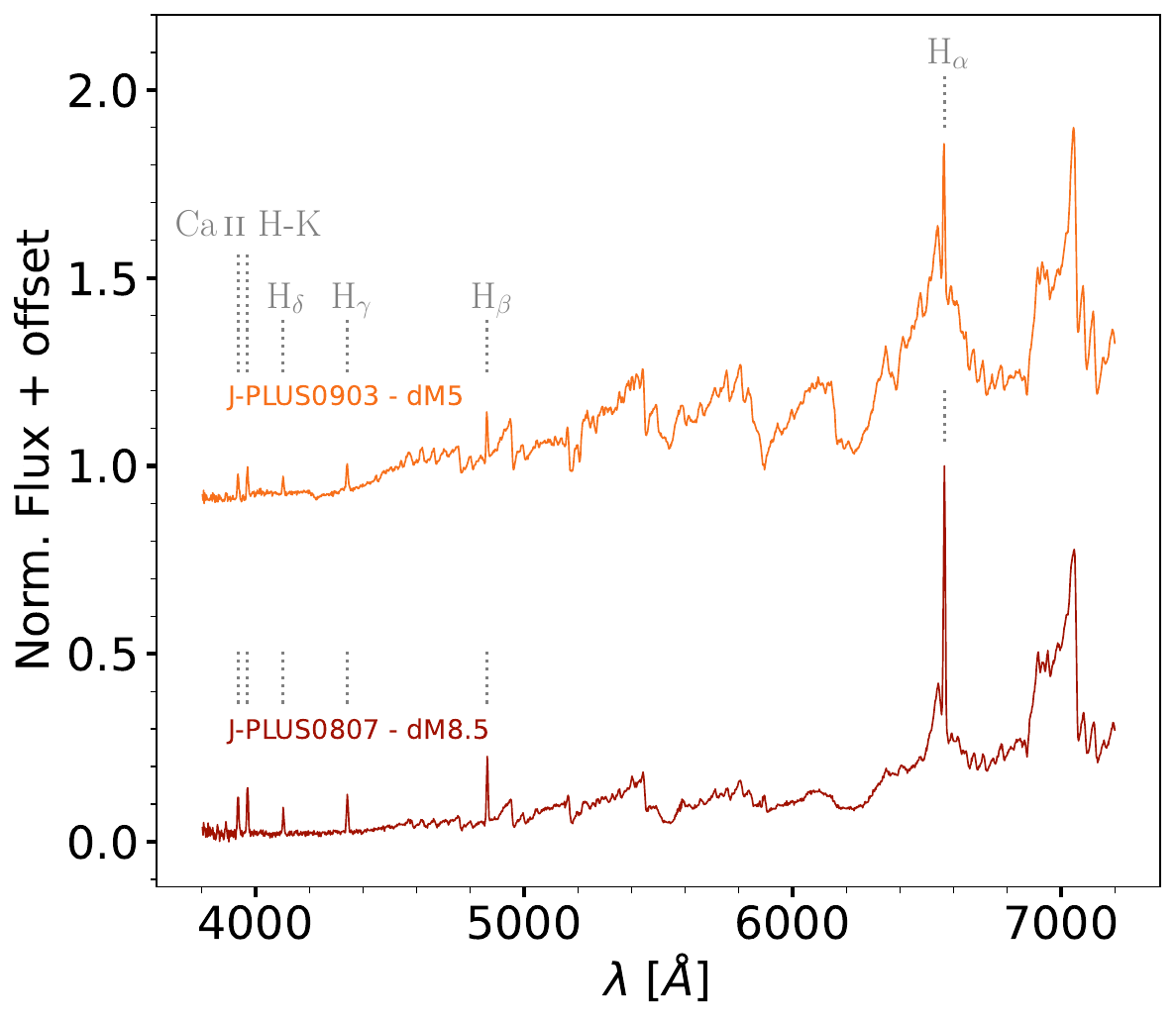}
    \caption{Zoomed-in view of the NOT/ALFOSC (\textit{left panel}) and GTC/OSIRIS (\textit{right panel}) co-added spectra of J-PLUS0807 and J-PLUS0903. The grey dashed lines mark Ca~{\sc ii} H and K, H$\delta$, H$\gamma$, H$\beta$, and H$\alpha$ emission lines.}
    \label{fig:spectra_red}
\end{figure*}

The obtained spectra confirm both J-PLUS0807 and J-PLUS0903 as active M dwarfs with Ca~{\sc ii} H and K and H$\alpha$ in emission, while the rest of the targets show no signs of activity. Figure \ref{fig:spectra_red} shows a close-up view of the spectral region of interest for these stars, with prominent Ca~{\sc ii} H and K, H$\delta$, H$\gamma$, H$\beta$ and H$\alpha$ emission lines. We quantified the H$\alpha$ emission using the \texttt{specutils}\footnote{\url{https://specutils.readthedocs.io/en/stable/index.html}} \citep{specutils} Python package, obtaining an H$\alpha$ equivalent width of $-16.80\,\AA$ and $-5.90\,\AA$ for the co-added NOT/ALFOSC spectra, and of $-18.36\,\AA$ and $-6.08\,\AA$ for the co-added GTC/OSIRIS spectra of J-PLUS0807 and J-PLUS0903, respectively. These results correspond to levels of H$\alpha$ emission that are not uncommon among this type of stars \citep{Schmidt2007,martin2010}. We find no significant differences between the equivalent width measurements for the individual spectra of each exposure. 

%--------------------------------------------------------------------

\subsection{Light curve analysis}
\label{sec:tess}

\noindent We queried the Mikulski Archive for Space Telescopes (MAST\footnote{\url{https://mast.stsci.edu/portal/Mashup/Clients/Mast/Portal.html}}) to fetch high-cadence photometric data for our sample. We found two-minute-cadence TESS-calibrated LCs for the two closest stars, J-PLUS0807 and J-PLUS0903, with TESS Input Catalog (TIC) IDs 461654150 and 166597074, respectively. Table \ref{tab:TESS_data} shows the details of the retrieved LCs, which are processed using the pipeline developed by the Science Processing Operations Centre \cite[SPOC;][]{spoc}. The contamination ratio, \texttt{Rcont}, listed in the TIC \citep{tic8_2} is 10\% and 0.22\% for J-PLUS0807 and J-PLUS0903, respectively. To further study a possible contamination of the TESS photometry for these two stars, we used the \texttt{tpfplotter}\footnote{\url{https://github.com/jlillo/tpfplotter}} \citep{tpfplotter} tool to explore the target pixel files (TPFs) of the fields of our targets. Thus, we only found a $\sim1\%$ contamination, obtained from the difference in \textit{Gaia} magnitudes, from \textit{Gaia} sources within the photometric apertures selected by the SPOC pipeline to process the LCs of the two stars.

\begin{table}
\fontsize{10pt}{10pt}\selectfont
 \caption{Details of two-minute-cadence TESS LCs used in this work.}
 \label{tab:TESS_data}
 \centering  
 \begin{tabular}{l c c}
  \hline\hline
  \noalign{\smallskip}
  
  TIC ID & TESS Sectors & Observation length\\
   & & [d]\\
  
  \noalign{\smallskip}
  \hline
  \noalign{\smallskip}
  
  461654150 & 20, 44, 45, 46, and 47 & 112.41 \\
  
  166597074 & 21 & 23.96 \\

  \noalign{\smallskip}
  \hline
 \end{tabular}
\end{table}

We identified clear evidence of flaring activity and periodic variability in the retrieved two-minute-cadence LCs for J-PLUS0807 and J-PLUS0903, which are analysed in detail in Sects. \ref{sec:flares} and \ref{sec:periods}. For the remaining five stars in our sample, which do not have processed, short-cadence TESS data, we used the Python package \texttt{lightkurve} \citep{lightkurve} to manually extract LCs from the TESS full frame images cutouts, but did not find any flare events or periodic variability signals. We also searched for time-resolved UV data from the NASA Galaxy Evolution Explorer \citep[GALEX;][]{galex} mission for our targets, using the \texttt{gPhoton} \citep{gphoton} database and software, but we did not find any.

%--------------------------------------------------------------------

\subsubsection{Flares}
\label{sec:flares}

\begin{figure*}
\centering
        \includegraphics[width=\linewidth]{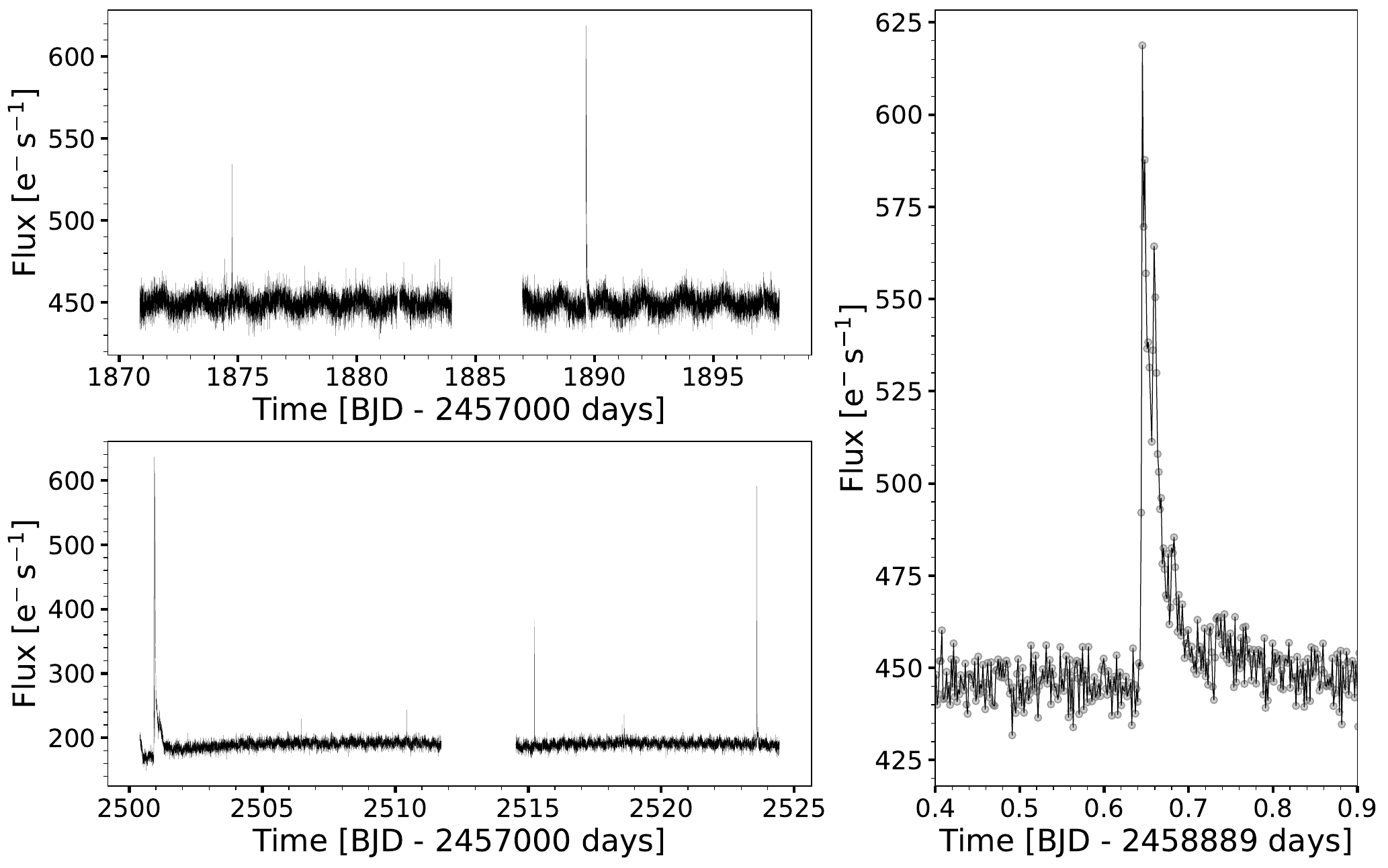}
    \caption{LCs of J-PLUS0903 (sector 21; \textit{top-left panel}) and J-PLUS0807 (sector 44; \textit{bottom-left panel}). The \textit{right panel} shows the largest flare event of the J-PLUS0903 LC.}
    \label{fig:flares_lc}
\end{figure*}

\noindent For our analysis, we used the Pre-search Data Conditioning Simple Aperture Photometry \citep[PDCSAP;][]{smith2012,stumpe2012,stumpe2014} flux, available in the TESS LCs retrieved for J-PLUS0807 and J-PLUS0903, which is already corrected from long-term trends, instrumental effects, and excess flux due to star-field crowding. We removed all data points with non-zero quality flags, after visually verifying that points with 512 (`impulsive outlier removed before co-trending') or 1024 (`cosmic ray detected on collateral pixel row or column') quality flags were not actually part of a real flare. We identified several flare events in all the J-PLUS0807 and J-PLUS0903 TESS LCs. For example, Fig. \ref{fig:flares_lc} shows the LC of J-PLUS0903 (top-left panel) and one of the LCs of J-PLUS807 (sector 44; bottom-left panel), with multiple flaring episodes observed in both of them. Moreover, the right panel provides a zoomed-in view of the flare event occurring around day 1890 (BJD 2457000 days) in the J-PLUS0903 LC.

We used the open-source Python software  \texttt{AltaiPony}\footnote{\url{https://altaipony.readthedocs.io/en/latest/}} \citep{davenport2016,ilin2021} to automatically identify and characterise flares in the LCs. Prior to flare detection, we detrended the LCs using a Savitzky-Golay filter \citep{Savitzky1964} to remove rotational modulation trends. For flare detection, we followed the same procedure as \citet{davenport2014}, \citet{doyle2019}, \citet{doyle2022}, and \citet{kumbhakar2023}, identifying flares as two or more consecutive points that are $2.5\sigma$ above the local scatter of the data \citep{chang2015}. As also reported by \citet{vida2017} and \citet{doyle2018}, we find no obvious relationship between flaring activity and rotational phase. \texttt{AltaiPony} automatically determines several flare properties, such as start and end times, flare amplitude, and equivalent duration, which is the area under the flare LC in units of seconds. Using the observed NOT/ALFOSC spectra and the tool \texttt{Specphot}\footnote{\url{http://svo2.cab.inta-csic.es/theory/specphot/}} \citep{specphot}, developed and maintained by the Spanish Virtual Observatory\footnote{\url{http://svo2.cab.inta-csic.es}}, we obtained the star quiescent flux in the TESS bandpass. We relied on the calculated flux and \textit{Gaia} distances of our targets to derive the quiescent stellar luminosity and multiplied it by the equivalent duration to obtain the flare energy in the TESS bandpass. We obtained $L_{\rm TESS}=2.3\times10^{29}$~erg\,s$^{-1}$ and $L_{\rm TESS}=2.2\times10^{30}$~erg\,s$^{-1}$ for the quiescent luminosity in the TESS bandpass of J-PLUS0807 and J-PLUS0903, respectively. Table \ref{tab:flares} details the flare properties for each target.

\begin{table*}
\fontsize{10pt}{10pt}\selectfont
 \caption{Detailed flare properties of our targets with processed TESS LCs.}
 \label{tab:flares}
 \centering          
 \begin{tabular}{l c c c c c}
  \hline\hline
  \noalign{\smallskip}
  
  Object & Sector & Number of flares & $\log(E)$ range & Duration range & Flare rate \\
   & & & [erg] & [min] & [d$^{-1}$]\\
  
  \noalign{\smallskip}
  \hline
  \noalign{\smallskip}
  
  J-PLUS0807 & 20 & 5 & 30.9--32.1 & 4.0--32.0 & 0.22\\
  
  J-PLUS0807 & 44 & 6 & 31.2--33.4 & 6.0--416.0 & 0.28\\
  
  J-PLUS0807 & 45 & 5 & 30.9--32.9 & 6.0--54.0 & 0.23\\

  J-PLUS0807 & 46 & 7 & 31.1--32.6 & 4.0--36.0 & 0.30\\

  J-PLUS0807 & 47 & 4 & 31.0--33.3 & 6.0--368.0 & 0.17\\

  J-PLUS0903 & 21 & 4 & 31.6--33.0 & 4.0--60.0 & 0.17\\

  \noalign{\smallskip}
  \hline
 \end{tabular}
\end{table*}
   
\begin{figure}
        \includegraphics[width=\columnwidth]{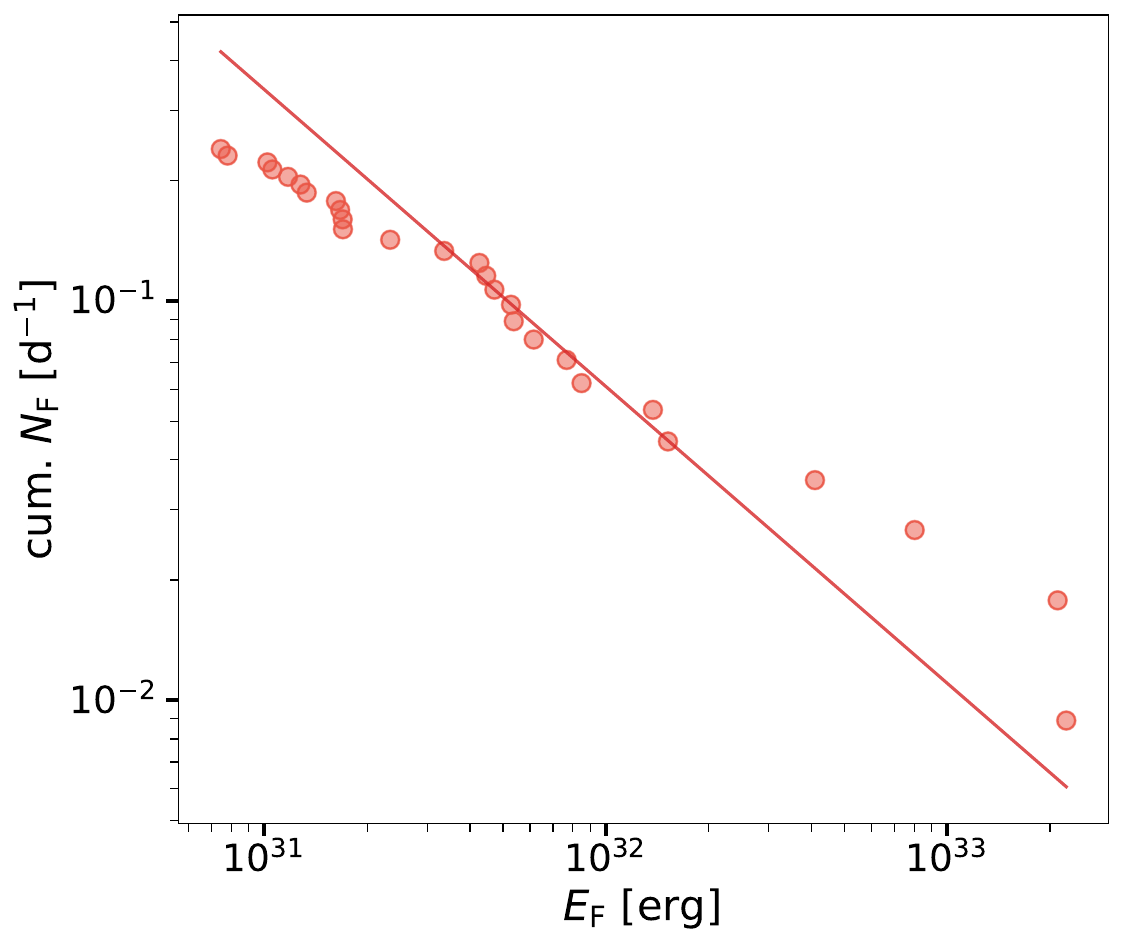}
    \caption{Cumulative FFD for J-PLUS0807. The solid red line represents the power-law fit obtained with \texttt{AltaiPony}.}
    \label{fig:ffd}
\end{figure}

The flare energy and rate obtained for our targets are typical of active, fast rotating mid- and late-M dwarfs \citep{doyle2019,ramsay2020,stelzer2022}. With the observed flares for J-PLUS0807, we built the cumulative flare frequency distribution (FFD) to study the flare rate as a function of flare energy. This was not possible for J-PLUS0903 due to the low number of events available. FFDs can be expressed as a power-law relation \citep{stelzer2007,lin2019}:

\begin{equation}    
    \frac{d \nu}{d E_{\rm F}}\sim E_{\rm F}^{-\alpha},
    \label{eq:ffd}
\end{equation}

\noindent where $\nu$ is the cumulative flare rate for a given flare energy $E_{\rm F}$, and $1-\alpha$ is the slope of a linear fit to a log-log representation. To fit the FFD, we relied on \texttt{AltaiPony}'s \texttt{FFD.fit\_powerlaw()} method, which fits the power-law parameters simultaneously using the Markov chain Monte Carlo method described in \citet{wheatland2004}. Since the detection probability decreases in the low-energy regime, where flares can go undetected due to the noise present in the LC, we discarded the low-energy tail of the FFD in the fit of the power-law \citep{hawley2014,chang2015,ilin2021}. Figure \ref{fig:ffd} shows how the power-law relation breaks down around $E_{\rm F}=10^{31.5}$~erg, which is the threshold we applied to consider flares in the FFD fitting. Following this methodology, we obtained $\alpha=1.74_{-0.17}^{+0.20}$ for J-PLUS0807, which is in agreement with what \citet{lin2019}, \citet{raetz2020}, and \citet{murray2022} found for their samples of 548, 56, and 85 flaring M dwarfs, respectively. We found that the less energetic flares, which are more frequent as illustrated in the FFD, are also shorter in duration and the easiest to detect with the J-PLUS observation strategy (see Sect. \ref{sec:sample}).

%--------------------------------------------------------------------

\subsubsection{Rotation periods}
\label{sec:periods}

\begin{figure*}
    \centering
        \includegraphics[width=8.5cm]{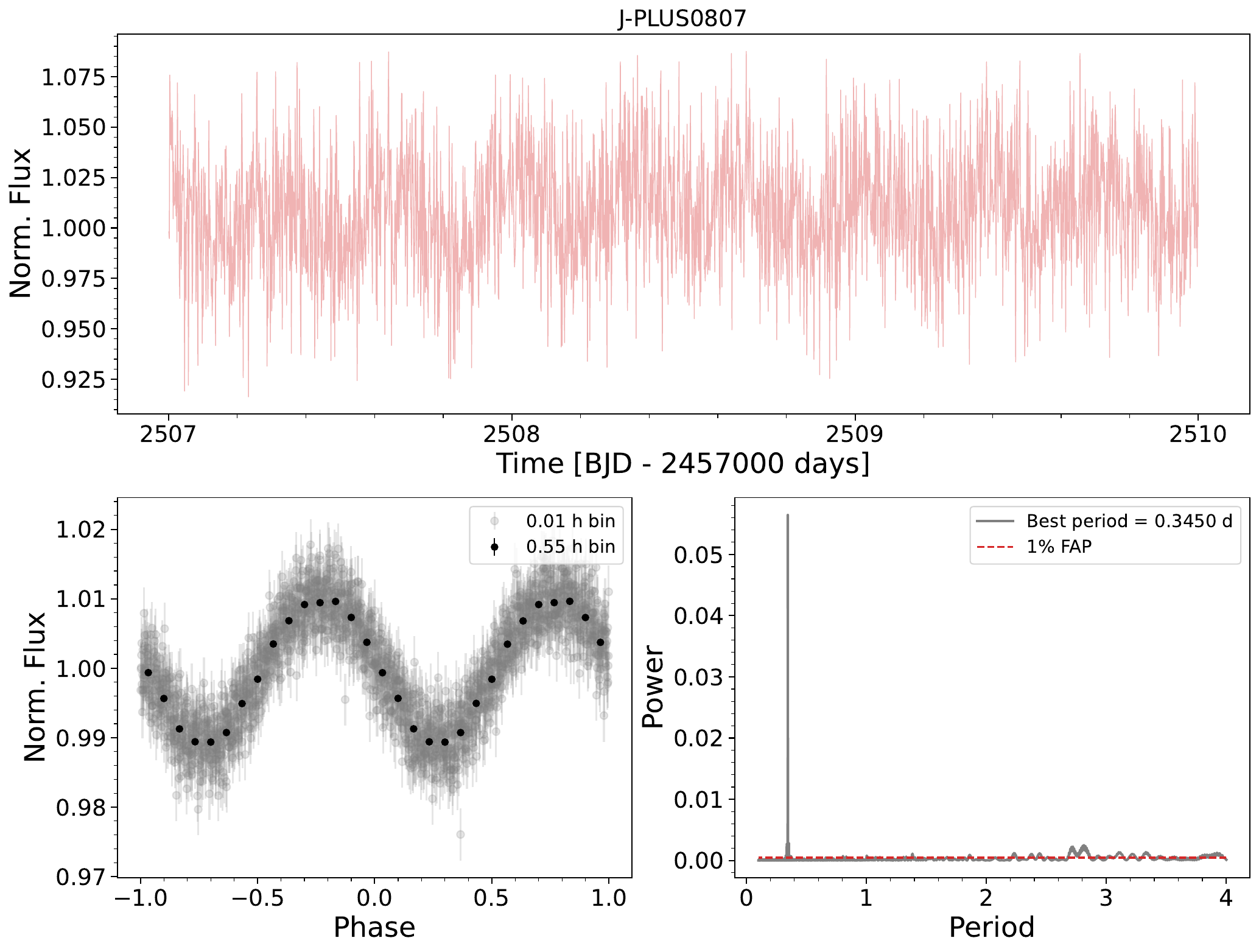}
            \includegraphics[width=8.5cm]{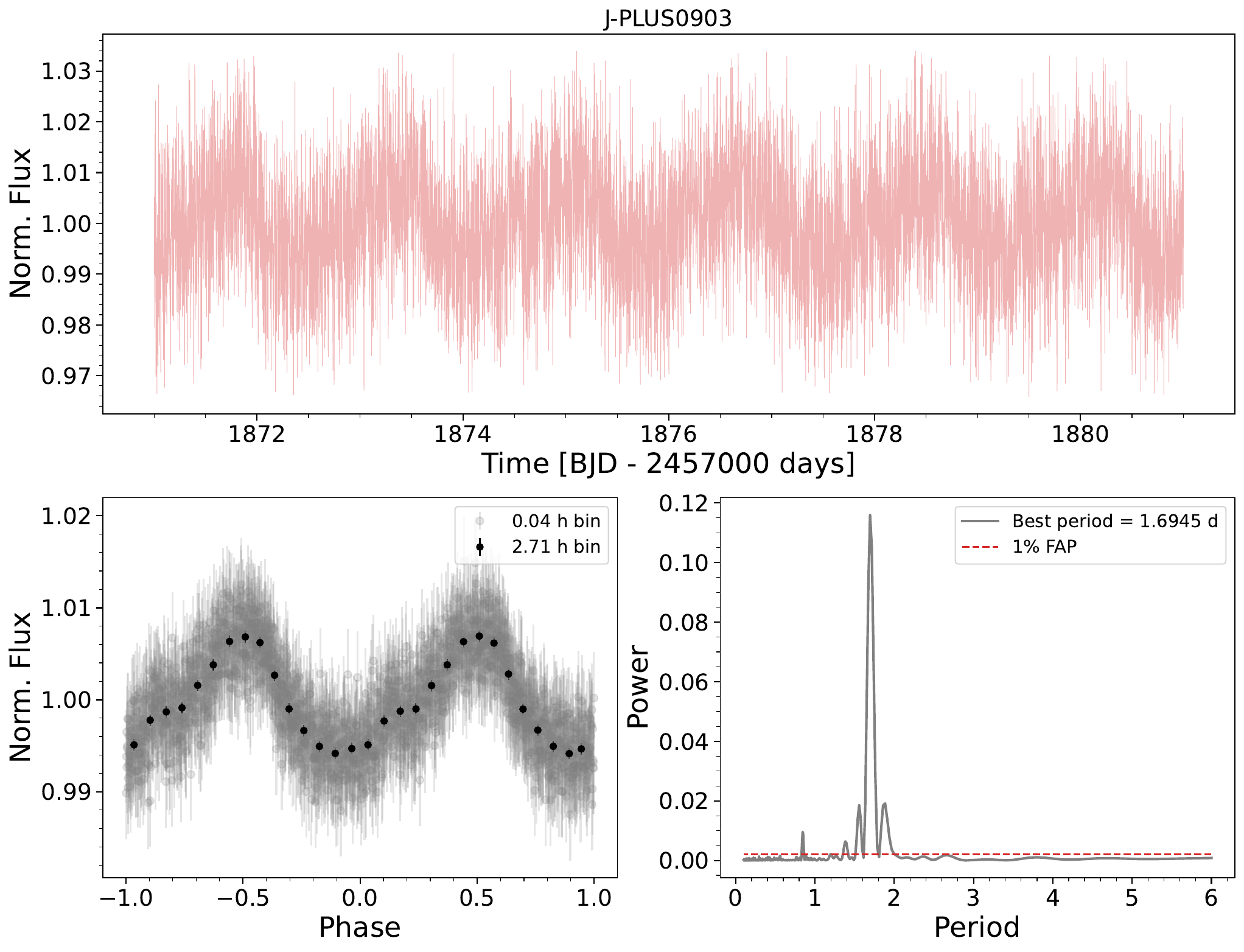}
    \caption{TESS LCs (\textit{top panels}), phase-folded LCs (\textit{bottom-left panels}), and periodograms (\textit{bottom-right panels}) for our two targets. The \textit{top panels} show only a small section of the LCs for better visibility. Two different bin sizes are shown for the binned phase-folded LCs, with grey and black dots. In the periodograms, the dashed red line represents the 1\% false alarm probability level.}
    \label{fig:periods}
\end{figure*}

\noindent All TESS LCs retrieved for J-PLUS0807 and J-PLUS0903 show a clear periodic variability, which usually arises due to co-rotating star spots that appear and disappear from the line of sight. Therefore, we relied on a Lomb-Scargle periodogram \citep{lomb,scargle}, using the \texttt{astropy} Python package \citep{astropy}, to search for the rotation period of each of our targets. Figure \ref{fig:periods} shows the periodogram for each of the  targets and the phase-folded LCs with the chosen periods, which are very prominent in the periodograms.

As a measure of the uncertainty of the peak position, we used the standard deviation of all the periods with a power greater than the half height of the periodogram peak. For J-PLUS0807, we computed the period using the data from all available sectors and obtained $P_{\rm rot}=0.3450\pm0.0002$\,d, which is consistent with the values reported by \citet{guangwei2024} using TESS data from sectors 20, 45, 46 and 47, \citet{seli2021} using only TESS data from sector 20, and \citet{newton2016}, who relied on photometry from the MEarth Project \citep{berta2012}. For J-PLUS0903, we obtained $P_{\rm rot}=1.69\pm0.02$\,d, which is the first estimation for the rotation period of this object. We confirmed that our targets were the sources of the detected variability using the \texttt{TESS\_localize}\footnote{\url{https://github.com/Higgins00/TESS-Localize}} \citep{tess_localize} Python package.

The computed rotation periods place our two targets as very fast rotators \citep{irwin2011}, which is deeply interlinked with the activity level observed. After reaching the main sequence, low-mass stars slowly spin-down due to the loss of angular momentum by stellar winds, and thus their magnetic activity decreases  over time \citep{yang2017,davenport2019,raetz2020}; this decrease may also be dependent on the stellar metallicity \citep{see24}, which makes obtaining robust age estimations for low-mass stars notoriously difficult. To explore this, we relied on \texttt{stardate}\footnote{\url{https://stardate.readthedocs.io/en/latest/}} \citep{stardate}, a Python tool that combines isochrone fitting with gyrochronology for measuring stellar ages. In our case, we included magnitudes from the Two-Micron All Sky Survey \citep[2MASS;][]{2MASS}, parallax and magnitudes from \textit{Gaia} DR3, magnitudes from the Sloan Digital Sky Survey \citep[SDSS;][]{sdss}, and the rotation periods obtained in this work as input parameters. Following this procedure, we obtained an age of $0.79^{+0.62}_{-0.09}$\,Gyr and $1.94^{+1.74}_{-1.26}$\,Gyr for J-PLUS0807 and J-PLUS0903, respectively, which is in agreement with the values found in the literature for fast rotators \citep{newton2016,doyle2019}. Here, the chosen value and uncertainties correspond to the median and $\pm1\sigma$ thresholds of the Markov chain Monte Carlo samples computed by \texttt{stardate}.

%--------------------------------------------------------------------

\section{Planetary habitability}
\label{sec:habitability}

\noindent Understanding the impact of the magnetic activity of M dwarfs on a planet's evolution and habitability is of crucial interest in the search for Earth-like planets. The common flaring activity and coronal mass ejections, together with the nearby habitable zone of these stars, can lead to substantial alteration of planetary atmospheres or even their erosion. It is unclear whether stellar flares are beneficial or detrimental to the habitability of exoplanets. It is possible that UV radiation emitted during flare events can trigger the development of prebiotic chemistry \citep{rimmer2018,airapetian2020}. Although abiogenesis would potentially be slower compared to prebiotic Earth due to the lower emission of M dwarfs at these wavelengths \citep{rugheimer2015,ranjan2017}, flares could provide the lacking UV energy \citep{buccino2007,jackman2023}. In this line, the flare events in Ca~{\sc ii} H and K emission lines revealed in this work may play an important role.

Continued exposure to $E_{\rm bol}>10^{34}$\,erg flares would make the presence of ozone layers impossible on any habitable zone terrestrial exoplanet orbiting an M dwarf \citep{tilley2019,chen2021}. Moreover, \citet{berger2024} recently demonstrated that the 9\,000\,K blackbody commonly assumed for flares underestimated the far-UV emission for 98\% of their sample, which would significantly increase the number of stars with sufficient flaring activity to fall into the ozone depletion zone from previous studies. Following the relation provided by \citet{seli2021}, we converted the TESS energies of the detected flares to bolometric flare energies. Thus, we obtained a rate of 0.02\,day$^{-1}$ for $E_{\rm bol}>10^{34}$\,erg flares for J-PLUS0807, which is an order of magnitude lower than the rate found by \citet{tilley2019} for the ozone layer to be eroded in habitable zone terrestrial exoplanets around M dwarfs. For J-PLUS0903, none of the flares exceeded this energy threshold.

%--------------------------------------------------------------------

\section{Conclusions} \label{sec:conclusions}

\noindent This work serves as a follow-up study of the sample of M dwarfs identified in our previous work \citep{masbuitrago22} with strong excess emission in the J-PLUS filters corresponding to Ca~{\sc ii} H and K and H$\alpha$ emission lines. Using low-resolution spectra collected with NOT/ALFOSC and GTC/OSIRIS, we measured the PC3 spectral index of our targets and spectroscopically confirmed the mid-M dwarf nature of six of them for the first time. We confirm that the strong excess emission detected in the J-PLUS photometry is caused by transient flare events, suggesting that common M dwarfs experience two types of flares, those already well known in H$\alpha$ and those in Ca~{\sc ii} H and K presented in this work. Work dedicated to the study of flares in large M dwarf samples usually focuses only on H$\alpha$ flare events, which could lead to an underestimation of the number of flaring M dwarfs. In the future, multi-wavelength simultaneous observations will be essential for further studying the flaring activity of M dwarfs.

We analysed two-minute-cadence TESS LCs for J-PLUS0807 and J-PLUS0903 and performed a thorough characterisation of the multiple flare events observed in them. We estimated the flare energies in the TESS bandpass and found them to be in the range $7.4\times10^{30}-2.2\times10^{33}$\,erg. We find clear signs of a periodic variability in the TESS LCs, confirming the previously reported ultra-fast rotating nature of J-PLUS0807 with data from sectors 20, 44, 45, 46, and 47. Also, we computed the rotation period of for J-PLUS0903 for the first time, $P_{\rm rot}=1.69\pm0.02$\,d.

This work demonstrates the potential of multi-filter photometric surveys such as J-PLUS or the upcoming J-PAS to systematically detect flare events in M dwarfs, especially episodes of strong Ca~{\sc ii} H and K line emission, which may have important implications for exoplanetary space weather and habitability studies. Using a detection algorithm such as the one developed in \citet{masbuitrago22}, it is possible to identify a sample of candidates that can be confirmed and analysed with spectroscopic follow-up and high-cadence photometric LCs from TESS or similar missions, such as K2. It also highlights the fundamental role of stellar flares in shaping the habitability of exoplanets. A high frequency of energetic flares implies that planets around these stars may experience significant atmospheric erosion and elevated levels of surface radiation, although it could also trigger the development of prebiotic chemistry.

%--------------------------------------------------------------------

\section*{Data availability}

\noindent All the resources, including files with the reduced spectra and processed TESS LCs, and the code to reproduce the figures displayed in Sect. \ref{sec:results} are publicly available at \texttt{GitHub}\footnote{\url{https://github.com/pedromasb/flaring-MDwarfs}}.

%--------------------------------------------------------------------

\begin{acknowledgements}

We acknowledge financial support from the Agencia Estatal de Investigaci\'on (AEI/10.13039/501100011033) of the Ministerio de Ciencia e Innovaci\'on and the European Union (ERC, SUBSTELLAR, project number 101054354) through projects PID2020-112949GB-I00 (Spanish Virtual Observatory \url{https://svo.cab.inta-csic.es}) and
PID2019-109522GB-C5[3], and the Instituto Nacional de T\'ecnica Aeroespacial through grant PRE-OVE.
Based on observations made with the Gran Telescopio Canarias (GTC), installed at the Spanish Observatorio del Roque de los Muchachos of the Instituto de Astrofísica de Canarias, on the island of La Palma. 
This work is (partly) based on data obtained with the instrument OSIRIS, built by a Consortium led by the Instituto de Astrofísica de Canarias in collaboration with the Instituto de Astronomía of the Universidad Autónoma de México. OSIRIS was funded by GRANTECAN and the National Plan of Astronomy and Astrophysics of the Spanish Government.
Based on observations made with the Nordic Optical Telescope, owned in collaboration by the University of Turku and Aarhus University, and operated jointly by Aarhus University, the University of Turku and the University of Oslo, representing Denmark, Finland and Norway, the University of Iceland and Stockholm University at the Observatorio del Roque de los Muchachos, La Palma, Spain, of the Instituto de Astrofisica de Canarias. 
The data presented here were obtained [in part] with ALFOSC, which is provided by the Instituto de Astrofisica de Andalucia (IAA) under a joint agreement with the University of Copenhagen and NOT.
We thank NOT support astronomers David Jones and Tapio Pursimo for their help on site.
This publication made use of VOSA, developed under the Spanish Virtual Observatory (https://svo.cab.inta-csic.es) project funded by MCIN/AEI/10.13039/501100011033/ through grant PID2020-112949GB-I00.
This work made use of \texttt{tpfplotter} by J. Lillo-Box (publicly available in www.github.com/jlillo/tpfplotter).
We made extensive use of Python throughout the entire process, including the packages \texttt{pandas} (\url{https://github.com/pandas-dev/pandas}, \texttt{numpy} \citep{numpy}, \texttt{matplotlib} \citep{matplotlib}, \texttt{astropy} \citep{astropy}, and \texttt{lightkurve}, a Python package for Kepler and TESS data analysis \citep{lightkurve}.

\end{acknowledgements}

\bibliographystyle{aa} % style aa.bst
\bibliography{Bibliography} % your references Yourfile.bib

\appendix
\section{Additional figures}
\label{app_a}

\begin{figure}[h!]
        \includegraphics[width=.75\columnwidth]{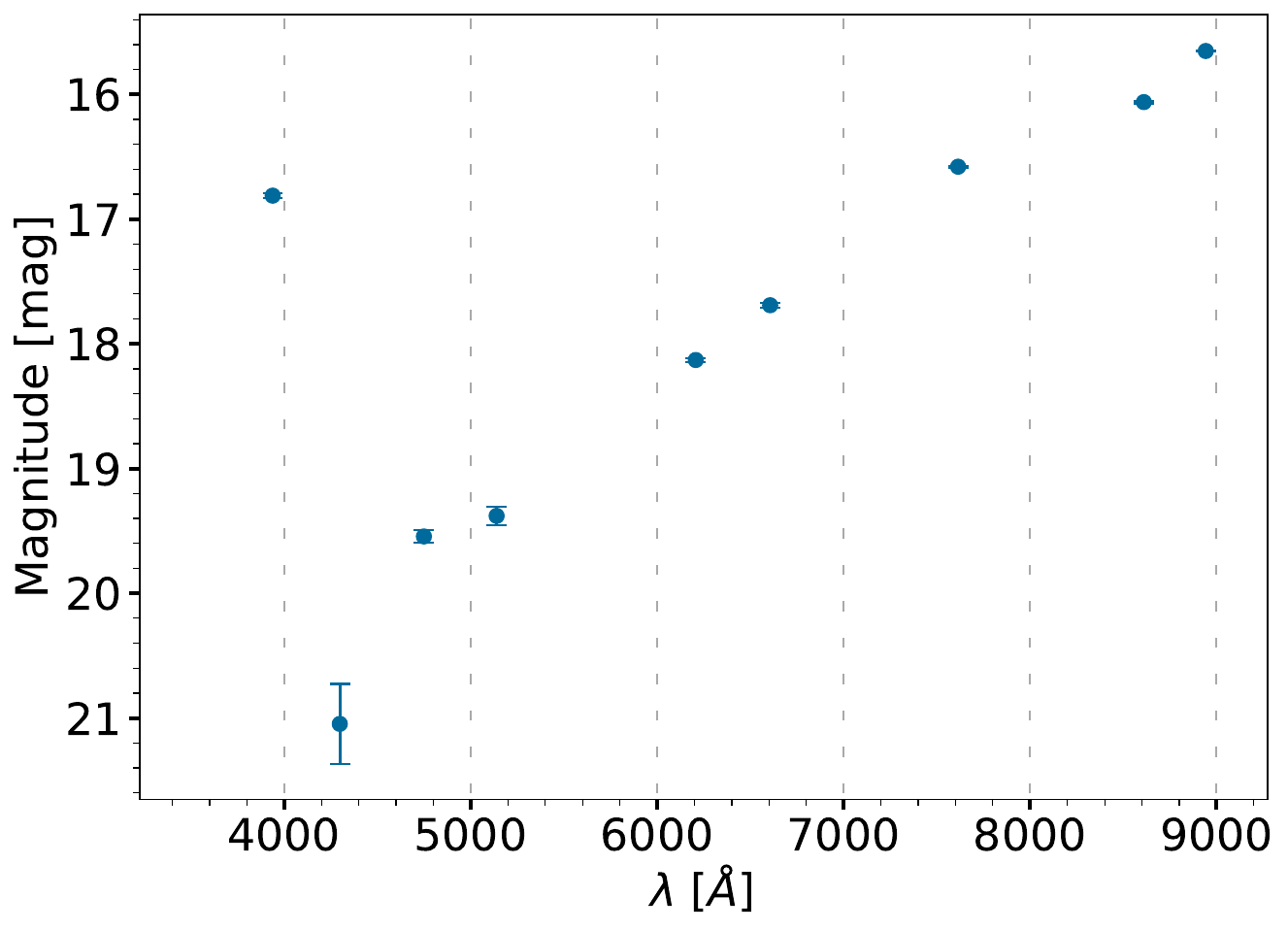}
    \caption{J-PLUS photometry for J-PLUS0114.}
    \label{fig:jplus_0114}
\end{figure}

\begin{figure}[h!]
        \includegraphics[width=.75\columnwidth]{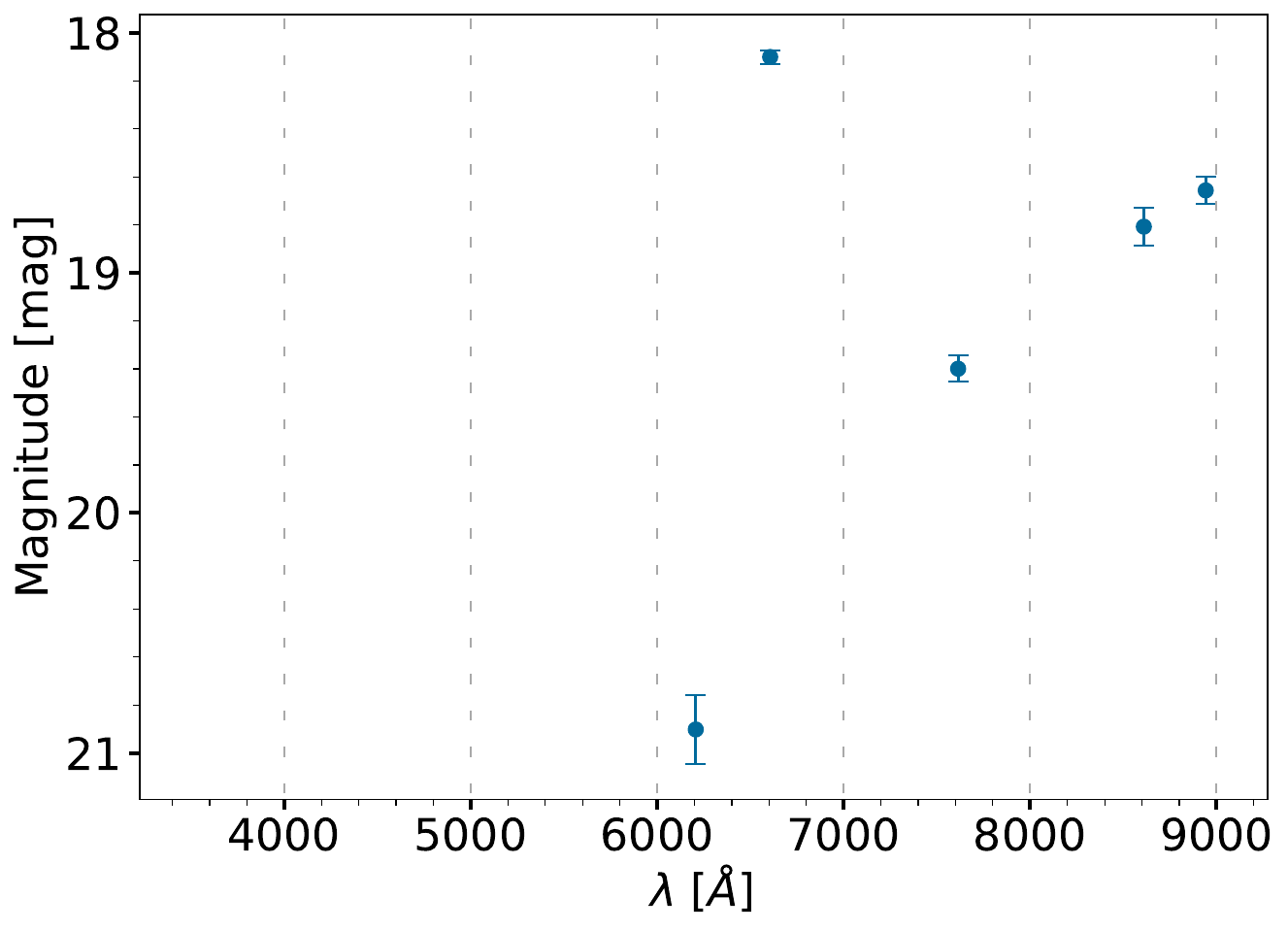}
    \caption{J-PLUS photometry for J-PLUS0226.}
    \label{fig:jplus_0226}
\end{figure}

\begin{figure}[h!]
        \includegraphics[width=.75\columnwidth]{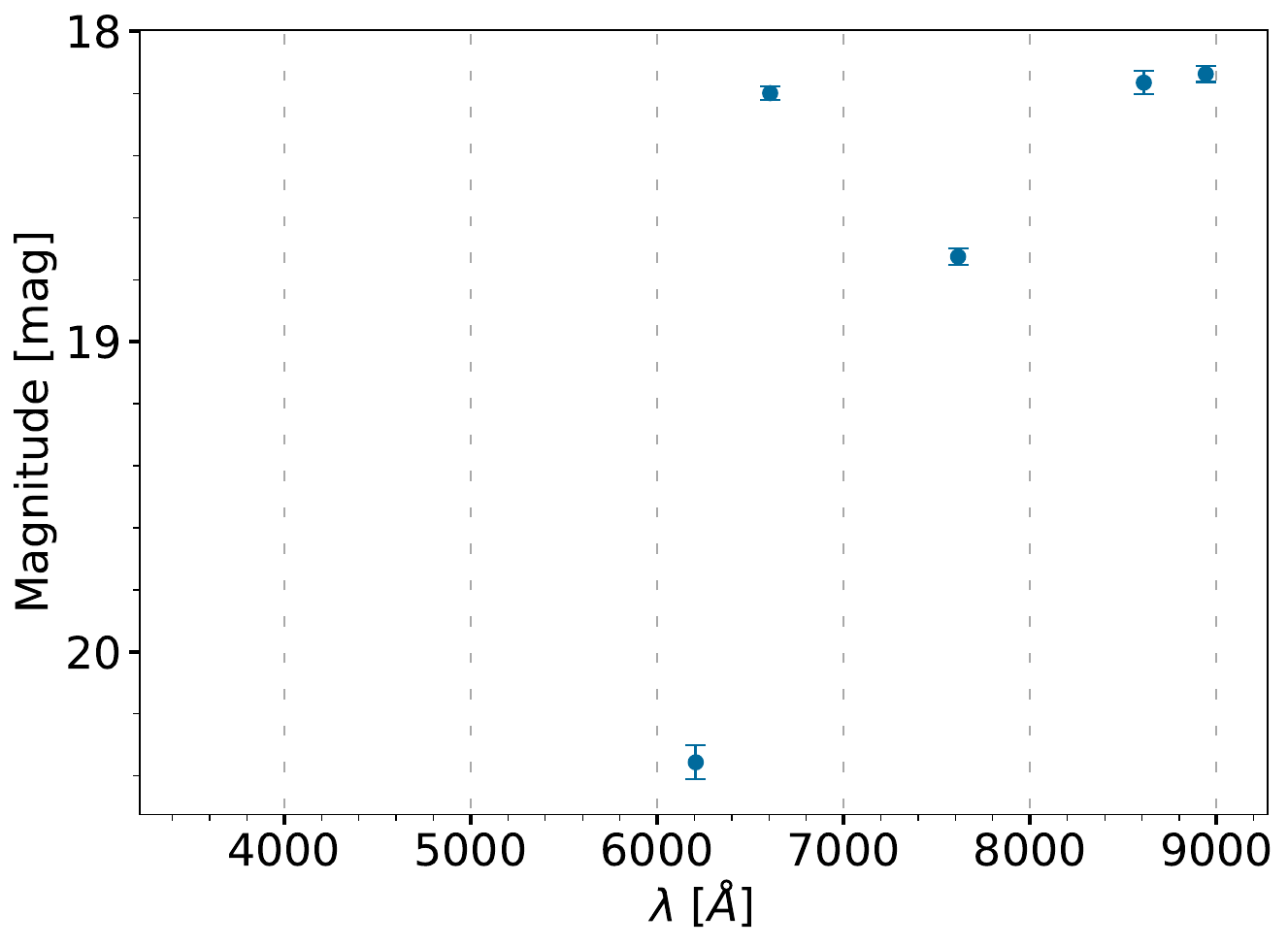}
    \caption{J-PLUS photometry for J-PLUS0708.}
    \label{fig:jplus_0708}
\end{figure}

\begin{figure}[h!]
        \includegraphics[width=.75\columnwidth]{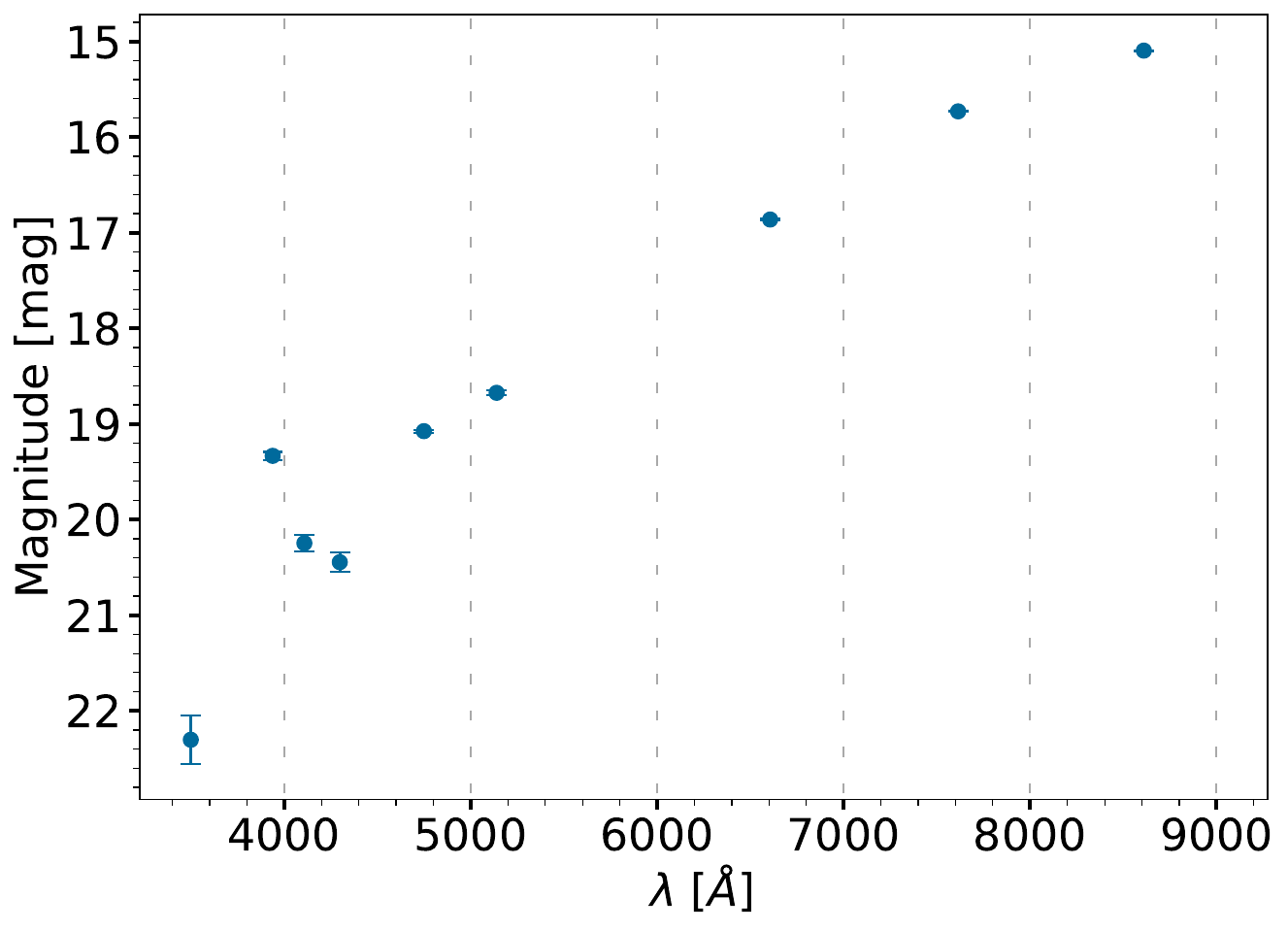}
    \caption{J-PLUS photometry for J-PLUS0744.}
    \label{fig:jplus_0744}
\end{figure}

\begin{figure}[h!]
        \includegraphics[width=.75\columnwidth]{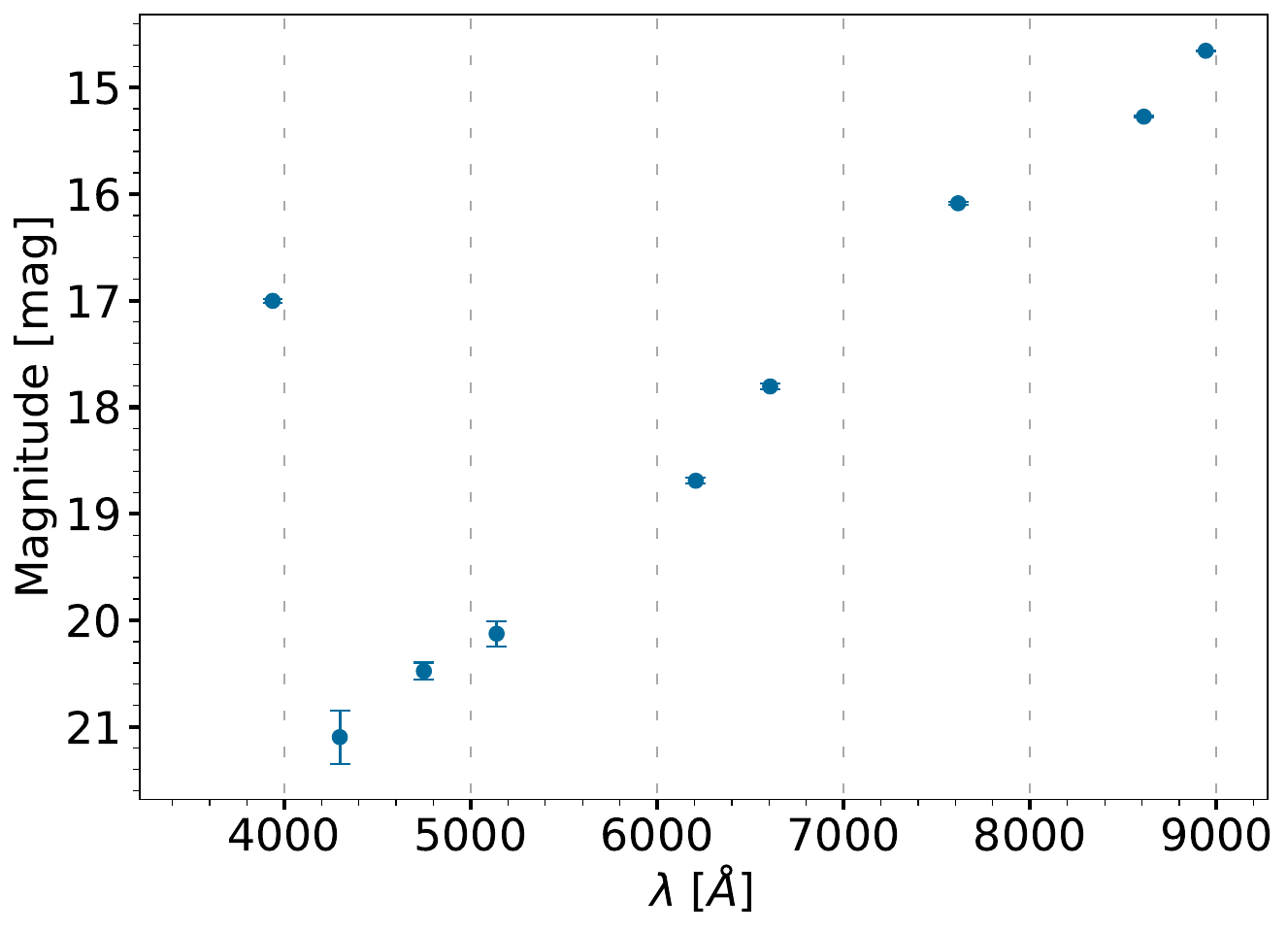}
    \caption{J-PLUS photometry for J-PLUS0807.}
    \label{fig:jplus_0807}
\end{figure}

\begin{figure}[h!]
        \includegraphics[width=.75\columnwidth]{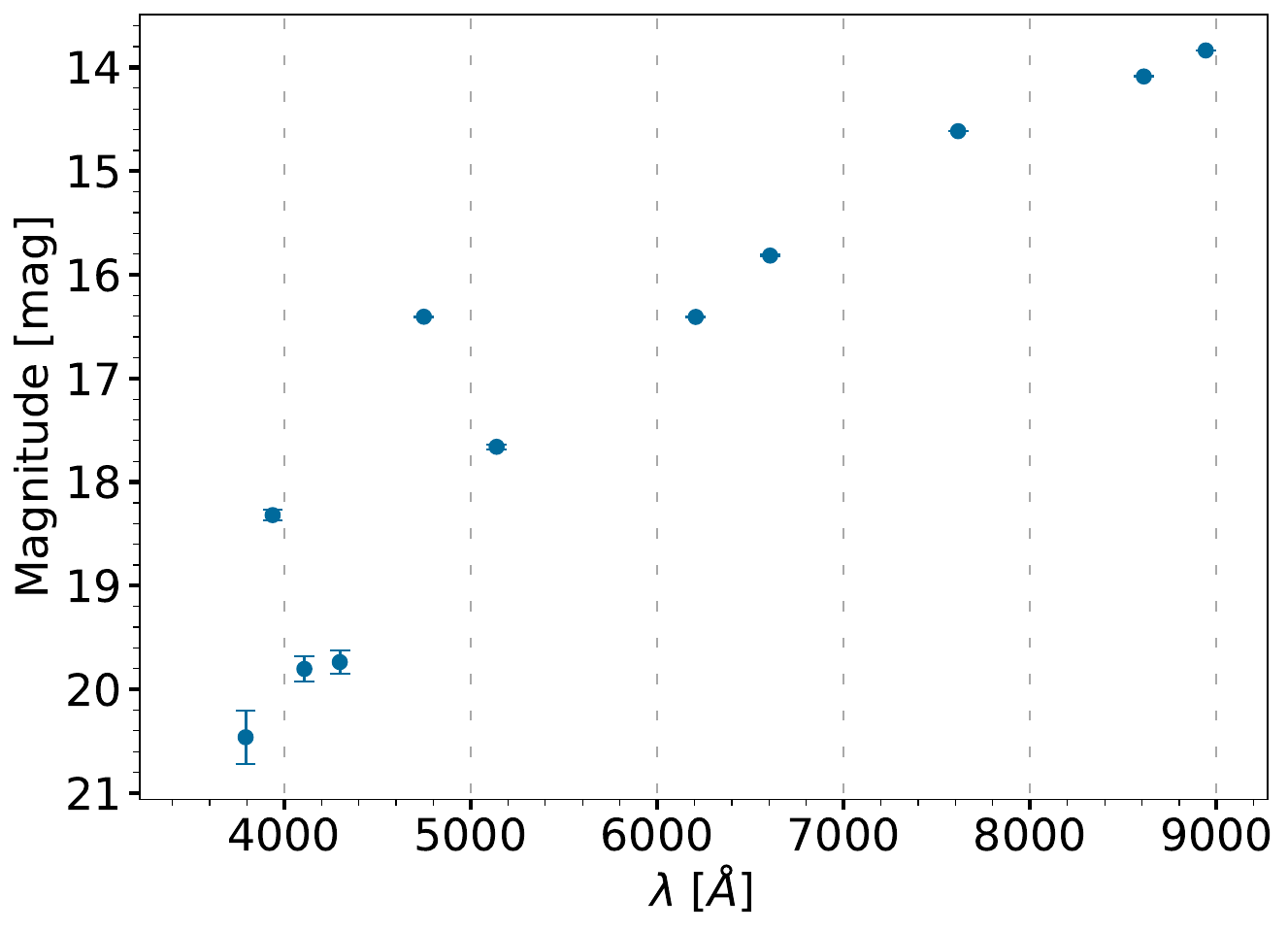}
    \caption{J-PLUS photometry for J-PLUS0903.}
    \label{fig:jplus_0903}
\end{figure}

\begin{figure}[h!]
        \includegraphics[width=.75\columnwidth]{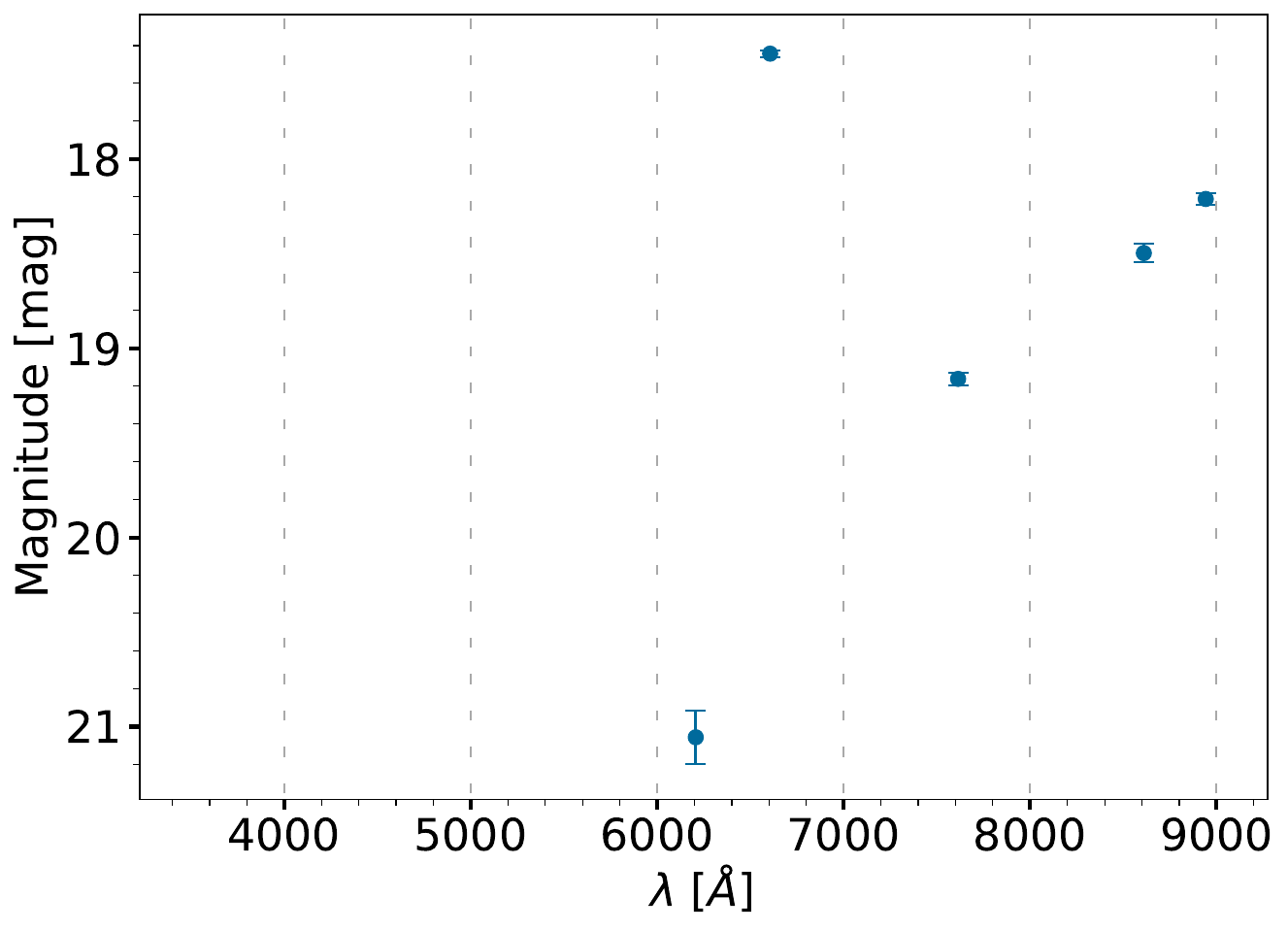}
    \caption{J-PLUS photometry for J-PLUS0914.}
    \label{fig:jplus_0914}
\end{figure}

\noindent In this appendix we provide the J-PLUS SEDs of each target star, as discussed in Sect. \ref{sec:reduced_sp}. Figures \ref{fig:jplus_0114}, \ref{fig:jplus_0226}, \ref{fig:jplus_0708}, \ref{fig:jplus_0744}, \ref{fig:jplus_0807}, \ref{fig:jplus_0903}, and \ref{fig:jplus_0914} show the SED of J-PLUS0114, J-PLUS0226, J-PLUS0708, J-PLUS0744, J-PLUS0807, J-PLUS0903, and J-PLUS0914, respectively.

% ---------------------------------------------------------------------------

\onecolumn
\section{Record of observations}
\label{app_b}

\begin{table*}[h!]
 \caption{Low-resolution spectra collected with NOT/ALFOSC and GTC/OSIRIS.}
 \label{tab:obs}
 \centering          
 \begin{tabular}{l c c c c c c c}
  \hline\hline
  \noalign{\smallskip}
  
  Object  &  Date  &  Configuration & Slit width & Grism & Exposures\\
  
  \noalign{\smallskip}
  \hline
  \noalign{\smallskip}
      
    J-PLUS0114 & 26 Jan. 2023 & ALFOSC Long Slit & 1.0" & \#4 & 1800s$\times$2\\
    
    J-PLUS0114 & 27 Jan. 2023 & ALFOSC Long Slit & 1.0" & \#4 & 1800s$\times$1\\
    
    J-PLUS0226 & 27 Jan. 2023 & ALFOSC Long Slit & 1.0" & \#4 & 2000s$\times$5\\
    
    J-PLUS0708 & 26 Jan. 2023 & ALFOSC Long Slit & 1.0" & \#4 & 2000s$\times$5\\
    
    J-PLUS0744 & 27 Jan. 2023 & ALFOSC Long Slit & 1.0" & \#4 & 1500s$\times$3\\
    
    J-PLUS0807 & 29 Oct. 2022 & OSIRIS Long Slit & 1.2" & R1000B & 180s$\times$6\\
    
    J-PLUS0807 & 26 Jan. 2023 & ALFOSC Long Slit & 1.0" & \#4 & 1800s$\times$2\\
    
    J-PLUS0807 & 27 Jan. 2023 & ALFOSC Long Slit & 1.0" & \#4 & 1800s$\times$1\\
    
    J-PLUS0903 & 29 Oct. 2022 & OSIRIS Long Slit & 1.2" & R1000B & 90s$\times$6\\

    J-PLUS0903 & 27 Jan. 2023 & ALFOSC Long Slit & 1.0" & \#4 & 1000s$\times$3\\
    
    J-PLUS0914 & 26 Jan. 2023 & ALFOSC Long Slit & 1.0" & \#4 & 2000s$\times$4\\
    
    J-PLUS0914 & 27 Jan. 2023 & ALFOSC Long Slit & 1.0" & \#4 & 2000s$\times$1\\

  \noalign{\smallskip}
  \hline
 \end{tabular}
 \tablefoot{Targets are ordered first by right ascension and then observation date.}
\end{table*}

\end{document}